# Possible routes to superconductivity in the surface layers of V-doped $Mg_{1-\delta}Ti_2O_4$ through multiple charge transfers and suppression of Jahn-Teller activity

Dibyendu Dey[1,*], T. Maitra[2], and A. Taraphder[3]

[1]Department of Physics and Astronomy, University of Maine, Orono, Maine 04469, USA

[2]Department of Physics, Indian Institute of Technology Roorkee, Roorkee, Uttarakhand 247667, India

[3]Department of Physics, Indian Institute of Technology Kharagpur, Kharagpur 721302, India

[*]Email: dibyendu.dey@maine.edu

**Abstract**

Superconductivity in the family of spinel oxides is very rare owing to their robust Mott-insulating nature. About half a century ago, $LiTi_2O_4$ became the first reported spinel compound to show superconductivity with a 12K transition temperature. Since then, several unsuccessful attempts were made to enhance the $T_c$ of this family of materials. However, a very recent experiment [arXiv:2209.02053] has reported superconductivity at a higher temperature (below 16K) in the V-doped $Mg_{1-\delta}Ti_2O_4$ thin surface layer while its bulk counterpart remains Mott insulating. The superconducting $T_c$ of this material is significantly higher compared to other engineered $MgTi_2O_4$ thin films grown on different substrates. From our first-principles analysis, we have identified that Mg-depletion significantly reduces Jahn-Teller (JT) activity and antiferromagnetic superexchange at the surface layer of V-doped $Mg_{1-\delta}Ti_2O_4$ due to considerable charge transfer between various ions. The combined effect of a degraded antiferromagnetic order and reduced JT activity weakens the Mottness of the system, leading to the emergence of superconductivity at higher temperatures.

**Introduction**

The emergence of unconventional superconductivity in strongly correlated systems is believed to originate from a repulsive electron-electron interaction that mediates the Cooper pair formation in the doped Mott insulators [1–4]. This strong on-site Coulomb interaction between electrons also drives the long-range anti-ferromagnetic order in the undoped compounds [5,6]. However, these two phases do not coexist and superconductivity arises after the suppression of anti-ferromagnetism and other intertwined electronic phases, e.g., possible spin and charge order [7,8]. Over the last four decades, unconventional superconductivity has been discovered in many strongly correlated systems: among them, Cuprates [9–12], Nickelates [13–15], iron-based pnictides [16–18], and heavy fermion superconductors [19–21] have been extensively studied.. Lately, the focus has shifted towards exploring new superconductors that offer rich functionalities. This family of superconductors not only helps in understanding the interplay between superconductivity and other exotic phases but also promises diverse application possibilities [22–25].

Transition metal spinel oxides show some promise as they are well known for a variety of fascinating functional properties due to a strong coupling between their charge, spin, orbital, and lattice degrees of freedom induced partly by Jahn-Teller (JT) distortion [26–31]. Charge fluctuations have a major role to play in enhanced superconductivity. Indeed, one of the motivations for the early search for high Tc superconductivity was the suppression of JT distortions in perovskite oxides leading to enhanced charge fluctuations [32]. However, superconductivity in the family of spinel oxides is rare. It was first observed in charge-frustrated mixed valent spinel oxide $LiTi_2O_4$ (having both $Ti^{3+}$: $3d^1$ and $Ti^{4+}$: $3d^0$) a long time back [33] with a $T_c$ of 12K. Since then, there has been a lively debate on the origin and nature of superconductivity in $LiTi_2O_4$, until a very recent experiment provided strong evidence for anisotropic *d*-wave pairing in $LiTi_2O_4$ [34]. This indeed suggests an unconventional superconductivity due to spin-orbital fluctuations induced by charge frustration [35,36].

Several attempts have been made in the past to increase the $T_c$ by doping at the Li/Ti site of $LiTi_2O_4$ [37–39] or finding alternate spinel oxides that have similar properties to $LiTi_2O_4$. In this regard, $MgTi_2O_4$ is one of the best candidates because the ionic radii of $Mg^{2+}$ and $Li^+$ are almost identical. However, like most spinel compounds, bulk $MgTi_2O_4$ undergoes a metal-to-insulator transition (driven by JT distortion) at 260 K [26,40] and remains a Mott insulator down to the lowest temperatures. Interestingly, a few purposefully engineered $MgTi_2O_4$ compounds bring superconductivity into the picture by suppressing its Mott insulating nature. For example, superconductivity has been reported in a superlattice consisting of $MgTi_2O_4$ and $SrTiO_3$ having a reduced Mg/Ti ratio [23] or in highly Mg-deficient Mg-Ti-O films (Mg: $Ti_9O_{10}$) on the (011)-oriented $MgAl_2O_4$ substrate [41]. Unfortunately, in all the above cases, the superconducting $T_c$ remains much lower than 12K, the $T_c$ of $LiTi_2O_4$ [23,33,41]. A major breakthrough came very recently when a recent experiment reported superconductivity in the Mg deficient surface layer of V-doped $MgTi_2O_4$ with a considerably higher $T_c$ of 16 K (the highest among the spinels so far) [42]. The bulk V-doped $MgTi_2O_4$ compound, on the other hand, remains a Mott insulator [43,44]. Despite the fact that charge and orbital fluctuations have been speculated as possible reasons for the enhancement of $T_c$, the mechanism and a theoretical explanation for superconductivity in the surface layer of the V-doped $Mg_{1-\delta}Ti_2O_4$ are still missing and require further attention.

Using first-principles approach, we have studied the electronic properties of insulating bulk V-doped $MgTi_2O_4$ and its Mg-deficient surface layer, which shows superconductivity at 16 K. We have demonstrated how various ordered phases such as spin, orbital, and charge-ordered states vary from bulk to surface layers, and how Mg-depletion at the top layer diminishes the overall JT activity of the neighboring layers and enhance the instabilities that already exist in the system. All these factors, as a whole, contribute to the 'high-Tc' superconductivity in V-doped $Mg_{1-\delta}Ti_2O_4$ spinel.

**Methods**

Density functional theory (DFT) calculations have been performed using a plane-wave basis with a kinetic energy cutoff of 500 eV and the projector augmented-wave (PAW) method [45,46] as implemented in the Vienna Ab-initio Simulation Package (VASP) [47,48]. The Perdew-Burke-Ernzerhof (PBE) [49] version of the generalized gradient approximation (GGA) has been used as an exchange-correlation functional. A $\sqrt{2} \times \sqrt{2} \times 1$ supercell containing 8 f.u. has been used for bulk calculations. This supercell serves two purposes - i) it can accommodate 37.5% bulk and 33.33% surface V doping (close to experimental 30% V-doped case) and ii) the up-up-down-down antiferromagnetic spin configuration (magnetic ground state) can be constructed. In spin-polarized calculations, strong correlation effects for the Ti and V 3d electrons have been incorporated within the GGA+$U$ approach [50]. We have used an effective on-site Coulomb repulsion, $U=$ 3 eV for both transition metal ions, which is reasonable for similar systems [51,52]. For the bulk compound, both lattice constants and atomic positions are relaxed using GGA+$U$ until the forces on each atom converged to 0.005 eV/Å. However, for the slab calculations, only atomic positions were relaxed until the forces on each atom converged to 0.01 eV/Å. The optimized bulk and slab structures are given in supplemental material S1. The reciprocal space integration has been carried out using 4×4×4 and 2×2×1 $\Gamma$-centered k-mesh grid for the bulk and slab calculations, respectively.

**Results**

Bulk V-doped $MgTi_2O_4$ is reported to crystallize in the distorted cubic-spinel structure (space group Fd-3m) above room temperature, whereas the XRD spectrum of V-doped $MgTi_2O_4$ at 15 K fits a tetragonal $I4_1/a$ structure better [44], indicating that the system undergoes a JT driven structural phase transition. Besides, it has been reported that the electronic structures and orbital order qualitatively remain almost identical irrespective of the choice of crystal structure [44]. So, either structure would do equally well while exploring electronic properties, but since the 'high-T$_c$' superconductivity has been observed below 16 K, we consider the tetragonal structure in our calculations.

For both the end compounds $MgTi_2O_4$ and $MgV_2O_4$, the respective transition metal $Ti^{3+}$ ($3d^1$) and $V^{3+}$ ($3d^2$) ions are JT active because of the partial occupation of t$_{2g}$ orbitals; they possess orbital degrees of freedom. However, when a $V^{3+}$ ion is doped at the $Ti^{3+}$ site, an electron is transferred from the Ti ion to the nearest V ion, and this introduces mixed valency for both V and Ti ions into the system. Therefore, V-doped $MgTi_2O_4$ contains both orbitally inactive $V^{2+}$($3d^3$) and $Ti^{4+}$($3d^0$) along with orbitally active $Ti^{3+}$ and $V^{3+}$ ions (Fig. 1a). Therefore, in V-doped MgTi$_2$O$_4$, the presence of Ti$^{4+}$ and V$^{2+}$ ions leads to a partial reduction of the orbital degrees of freedom, since these ions are JT inactive due to their empty and half-filled t$_{2g}$ shells, respectively. As shown in Supplemental Figure S1a, TiO$_6$ and VO$_6$ octahedra in this compound share edges, suggesting the occurrence of a cooperative JT distortion. This effect is particularly pronounced when neighboring transition metals are also JT-active, as observed in compounds like MgTi$_2$O$_4$ and MgV$_2$O$_4$. However, in the case of V-doped MgTi$_2$O$_4$, the cooperative JT activity is reduced due to the presence of non-JT-active V$^{2+}$ and Ti$^{4+}$ sites.

We now discuss the electronic structure of the tetragonal phase of V-doped MgTi$_2$O$_4$. As reported in the literature, the bulk V-doped MgTi$_2$O$_4$ compound is a Mott insulator, in which Ti and V spins form an antiferromagnetic (AFM) order. To obtain an insulating ground state, GGA+$U$ calculations have been performed. As reported in the previous literature [44], for 37.5% V-doping, the atomic arrangements of Ti and V ions for which the total energy of the system is minimum has been chosen (Fig. 1a). The up-up-down-down AFM spin order within each corner-shared tetrahedral unit of transition metals is taken as the initial spin configuration. But, due to the electron transfer from Ti$^{3+}$ → V$^{3+}$, a few Ti$^{3+}$ ions convert to spinless Ti$^{4+}$ (d$^0$) ions, and the converged spin configuration of V-doped MgTi$_2$O$_4$ is shown in Fig 1b.

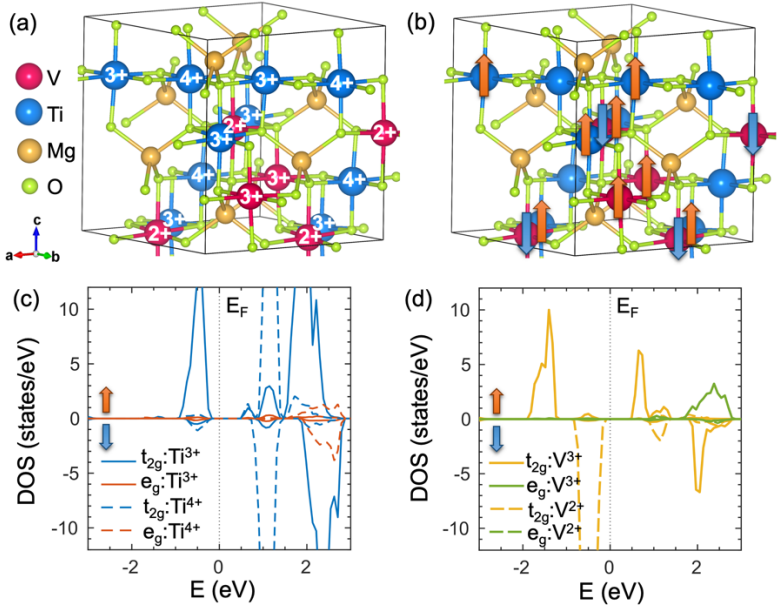

**Figure 1:** (a) Crystal structure and (b) magnetic ground state of 37.5% bulk V-doped MgTi$_2$O$_4$ compound containing both orbitally inactive Ti$^{4+}$(3d$^0$) and V$^{2+}$(3d$^3$) ions along with orbitally active Ti$^{3+}$ (3d$^1$) and V$^{3+}$ (3d$^2$) ions. Mg atoms are depicted in yellow, Ti are in blue, V are in red, and O are in green, whereas majority and minority spins are marked as orange and blue arrows, respectively. $d$-orbital (t$_{2g}$ and e$_g$) projected density of states (DOS) of (c) Ti and (d) V atoms in the majority and minority spin channels as obtained within GGA+$U$

calculations. The results indicate the insulating ground state of the 37.5% bulk V-doped MgTi$_2$O$_4$ compound, where the Fermi-level is denoted as $E_F$.

Figures 1c-d depict the resultant insulating density of states (DOS) of V-doped MgTi$_2$O$_4$. Here, Ti and V orbital-projected $d$-states (t$_{2g}$ and e$_g$) around the Fermi-level ($E_F$) are shown. As expected, the $d$-orbitals of Ti$^{4+}$ ions (d$^0$) are unoccupied in both spin channels, whereas the t$_{2g}$ levels of Ti$^{3+}$ (d$^1$) are partially occupied in the majority spin channel and completely unoccupied in the minority spin channel (Fig. 1c). On the other hand, the half-filled t$_{2g}$ levels of V$^{2+}$ (d$^3$) that lie well below the $E_F$ are occupied by the minority spins (Fig 1d). As a result, both Ti$^{4+}$ and V$^{2+}$ ions are orbitally inactive and do not participate in the JT activity. In case of V$^{3+}$ (d$^2$), t$_{2g}$ levels are partially occupied by two electrons in the majority spin channel (Fig. 1c). Interestingly, in the majority spin channel, the valance and conduction states (bands) across $E_F$ (~1 eV) are the t$_{2g}$ levels of Ti$^{3+}$ and V$^{3+}$, respectively. Therefore, under a suitable external perturbation (such as doping, electric field, etc.), the transfer of an electron from Ti$^{3+}$ → V$^{3+}$ is energetically more favorable than any other ions involved. Eventually, the electron-transfer process assists in lifting orbital degeneracies of electrons on both Ti and V ions and thereby melts the JT distortion. The incidental orbital order of Ti and V ions in the bulk V-doped MgTi$_2$O$_4$ compound is shown in Supplemental Fig. S1b. This partial DOS analysis to identify JT active/inactive sites is consistent with the length distortion parameter ($\zeta$) [53] analysis that involves Ti/V-O distortions of the respective octahedron (c.f. supplemental material S3 for details). But calculating $\zeta$ for the slab configurations is very challenging because of their complex geometry and the presence of many transition metal ions. Therefore, we have considered the partial DOS approach to identify JT active/non-active sites for bulk and slab structures.

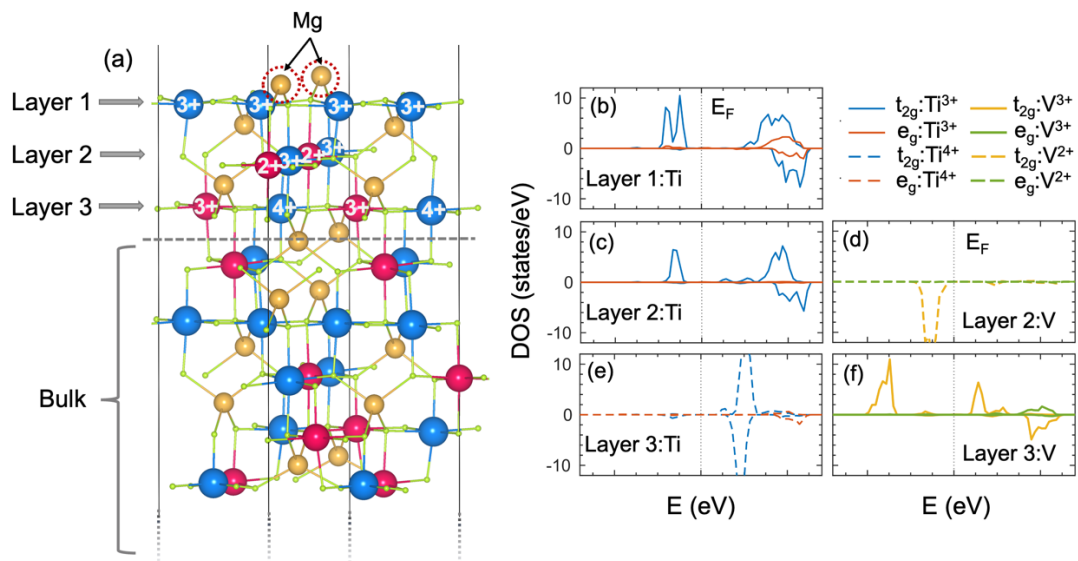

**Figure 2:** (a) V-doped MgTi$_2$O$_4$ slab structure containing 33.33% surface V atoms. The first three layers of transition metals are considered as the surface layers because of their proximity to the surface Mg atoms, whereas the bottom five layers can be considered as the part of bulk counterpart. The charge states of the respective transition metals of the surface layers are indicated. (b-f) *d*-orbital (t$_{2g}$ and e$_g$) projected density of states (DOS) of Ti and V atoms in the majority and minority spin channels of each surface layer as obtained within GGA+*U* calculations.

**Mg-depletion:** In order to understand the surface superconductivity phenomenon in the Mg-depleted V-doped MgTi$_2$O$_4$ compound, a slab has been constructed from the bulk compound. For this purpose, a $\sqrt{2} \times \sqrt{2} \times 2$ supercell has been constructed and a vacuum of 20 Å has been introduced along the (001) direction. The V-doped MgTi$_2$O$_4$ slab has the same stoichiometry as the bulk, and it contains 8 layers of transition metals. The advantage of this supercell construction is that the bulk magnetic configuration can be accurately reciprocated to the slab geometry, and the surface layers can be viewed separately from the bulk. The ratio of orbitally active V$^{3+}$ (Ti$^{3+}$) sites and orbitally inactive V$^{2+}$ (Ti$^{4+}$) sites in the V-doped MgTi$_2$O$_4$ slab remains the same 1:2 (3:2) as bulk. Therefore, the JT activity does not get perturbed by changing the dimensionality of the system.

First, we discuss the pristine V-doped MgTi$_2$O$_4$ slab (Fig. 2a) and analyze the DOS of each surface layer containing transition metal atoms. Figure 2b-h shows the *d*-orbital projected DOS of Ti and V atoms for the first three layers as these can be considered as the surface layers due to their proximity to the surface Mg atoms. The first layer contains only JT active Ti$^{3+}$ ions, and their t$_{2g}$ electrons occupy the states below the E$_F$ (Fig. 2b). The second layer, on the other hand, have both JT active Ti$^{3+}$ and JT inactive V$^{2+}$. The t$_{2g}$ electrons of the respective majority and minority spin channels of Ti$^{3+}$ and V$^{2+}$ lie below the E$_F$ (Fig. 2c-d). In the third layer, JT activity, however, is governed by V cations. As evident from the DOS (Fig. 2e), *d*-states of Ti atoms on the third layer are unoccupied, which confirms their +4 valency. Also, the DOS of V atoms (on the same layer) mimics the DOS of V$^{3+}$ in the bulk and shows the partial occupancy of t$_{2g}$ orbitals (Fig. 2f). Interestingly, these surface states remain insulating, which is consistent with the experiments. The corresponding orbital order of this slab structure is shown in Supplemental Fig. S2a.

Now we turn to discuss the case of Mg depleted V-doped MgTi$_2$O$_4$ slab structure (Fig. 3a), which causes a charge imbalance in the system. In order to restore the charge balance, a charge transfer process across the surface layers is inevitable. Besides, the vacancy defects created during this process would lead to local structural distortions that affect the JT activity and may favor superconductivity to appear in the system.

Therefore, understanding the electronic properties of the surface layer is crucial to explore the possibility of superconductivity at the surface of the V-doped $Mg_{1-\delta}Ti_2O_4$ spinel compound.

Figure 3b-f shows the DOS of Ti and V ions of the top three surface layers of V-doped $Mg_{1-\delta}Ti_2O_4$. Unlike the previous case (without Mg depletion), the first layer contains both JT active $Ti^{3+}$ and JT inactive $Ti^{4+}$ at alternate sites (cf. orbital order Supplemental Fig. S2b) as indicated in the DOS (Fig. 3b). The DOS of the next layer (Fig 3c), which was previously occupied by the $t_{2g}$ states of $Ti^{3+}$, becomes completely unoccupied indicating the transition of the $Ti^{3+} \rightarrow Ti^{4+}$. However, V atoms of the same layer remain in the +2 state (half-filled $t_{2g}$), which implies that the JT activity of the second layer is completely suppressed. Strikingly, the DOS of V atoms (Fig 3f) of the third layer shows a metallic nature ($t_{2g}$ levels lie on the $E_F$), and the ions show an intermediate valency close to 2+ that effectively reduces the strength of JT interaction at those sites. But the 4+ charge state of the Ti ion on the same layer does not change (Fig 3e). Note that charge states of both Ti and V ions in the fourth and subsequent layers remain unchanged suggesting that the effect of Mg-depletion on charge-transfer remains more prominent in the first three layers. Therefore, Mg depletion helps diminish the overall JT activity of the neighboring layers due to the emergence of more orbitally inactive Ti and V ions; and in addition, the slab becomes conducting.

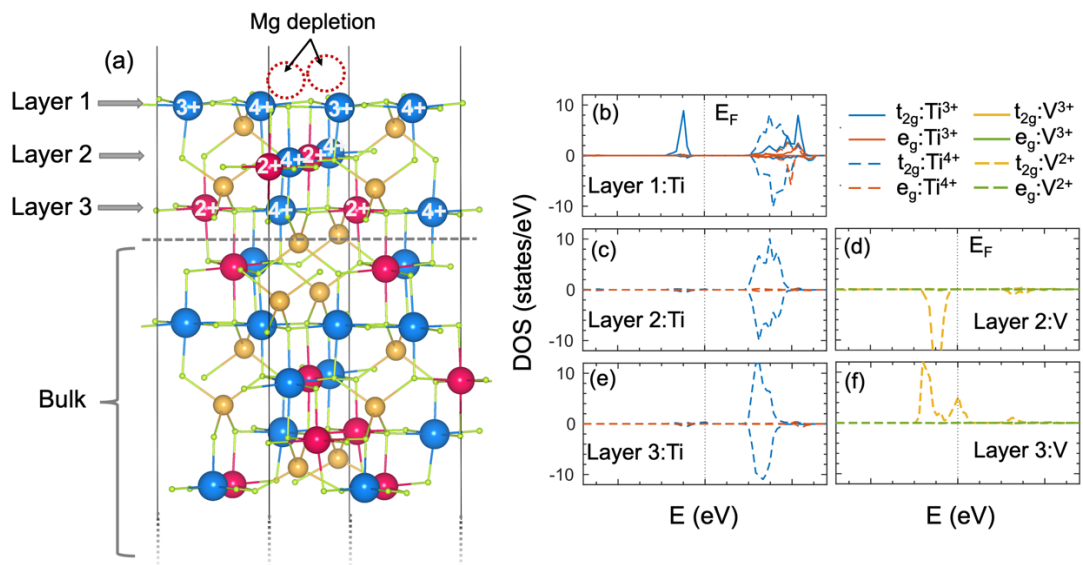

**Figure 3:** (a) V-doped $Mg_{1-\delta}Ti_2O_4$ (Mg-depleted) slab structure containing 33.33% surface V atoms. The charge states of the respective transition metals of the surface layers are indicated. (b-f) *d*-orbital ($t_{2g}$ and $e_g$) projected density of states (DOS) of Ti and V atoms in the majority and minority spin channels of each surface layer as obtained within GGA+*U* calculations.

Finally, we discuss the charge transfer process that happens at the surface layers of V-doped $Mg_{1-\delta}Ti_2O_4$. Figure 4a-b depicts the charge density difference between two slab structures before and after Mg depletion. The surface O ion, which is connected to both Mg and Ti receives electrons from Mg because the alkaline earth metal is more electro-positive than the transition metal. Therefore, when Mg is present at the surface, the unpaired valence electron of Ti stays at the d-orbital (yellow isosurface (Fig 4a): charge-surplus region). But as soon as the Mg is removed, Ti donates its electron to the O and becomes a charge-deficient site (cyan isosurface: Fig 4b). This electron transfer from Ti to O at the top layer explains the mechanism behind the transition of the Ti ion to the 4+ state. At the same time, another charge transfer process takes place from $Ti^{3+}$ (second layer) to $V^{3+}$ (third layer). The unpaired valence electron of $Ti^{3+}$ could not go to the $V^{2+}$ sites of the same layer as the $t_{2g}$ orbitals of the latter are half-filled and accommodating another electron would cost more energy. As a result, the electron would prefer to move to the $V^{3+}$ ion in the next layer to achieve the minimum energy configuration at both sites and become orbitally inactive simultaneously. The schematic diagram of the corresponding charge transfer process is shown in Fig. 4c.

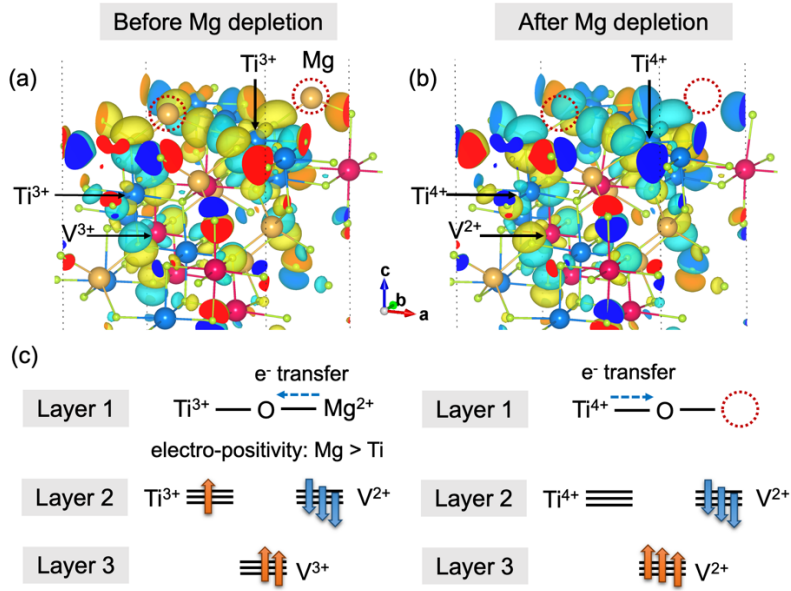

**Figure 4:** Charge density difference plot of the V-doped $MgTi_2O_4$ slab (a) with Mg and (b) without Mg atoms at the surface. The yellow and cyan colors represent the charge accumulation and charge depletion region, respectively. (c) A schematic of the charge transfer mechanism across the surface layers.

We have further analyzed the electronic structures and charge-states of 31.25% V-doped bulk $MgTi_2O_4$. 31.25% V-doped bulk compound has been constructed by replacing one V by Ti in 37.5% V-doped structure. This way, three different atomic arrangements of Ti and V ions are possible. These optimized structures and their associate energies have been included in the supplemental material S2 and supplemental

Table S1, respectively. Among all possible arrangements of Ti and V ions in this doping concentration, the structure shown in Fig. S3a has the lowest energy. The DOS of the respective Ti and V ions are shown in Supplemental Fig. S3b-c. The compound remains a Mott insulator, and the electronic structures remain qualitatively the same as before (37.5%). However, the slab constructed from 31.25% V-doped bulk has only 25% V atoms at the surface layer, below the experimental ~30% V-doped case [42]. The DOS (Supplemental Fig. S4) indicates that the insulating surface states of the slab remain gapped even after Mg-depletion. Also, the number of JT active ions are more on the surface layers compared to the previous case, where surface layers contain 33.33% V atoms. This finding indicates that 33.33% V ions at the surface layers could be an ideal situation for observing superconductivity at the V-doped $Mg_{1-\delta}Ti_2O_4$ thin films in agreement with the experiment.

**Discussion**

Mg depletion from the V-doped $MgTi_2O_4$ surface brings forth many possible scenarios that the observed surface superconductivity could be attributed to. First of all, charge and orbital fluctuations that were already present in the V-doped $MgTi_2O_4$ compound are further enhanced due to the removal of surface Mg atoms. This allows the system to undergo several electron-transfer processes that effectively stabilize energetically more favorable $Ti^{4+}$ and $V^{2+}$ states at the surface. Moreover, these charge states of the respective transition metals are JT inactive as well with reduced orbital degrees of freedom. In the recent literature [23,42], the suppressions of JT activity and orbital order were proposed as essential prerequisites for superconductivity. From our DFT calculations, we find this to be happening at the surface of V-doped $Mg_{1-\delta}Ti_2O_4$. Furthermore, the inclusion of $Ti^{4+}$ ions in the surface layers breaks the superexchange chain among neighboring ions due to their lack of magnetic moment ($d^0$), resulting in a weakening of the AFM order in the system, which in turn facilitates a superconducting transition.

**Conclusions**

In conclusion, through first-principles analysis, we have shown that the Mg depletion from the surface of V-doped $MgTi_2O_4$ increases the number of JT inactive sites significantly at the surface layers containing 33.33% V ions, which is lower than the nominal concentration of the bulk. The charge transfer process that happens across the surface layers to balance the charge deficiency (after the removal of $Mg^{2+}$ ions) has been identified as the possible driving mechanism for enhanced superconductivity. Besides, having more $Ti^{4+}$ states at the surface layers weakens the AFM superexchange interaction. Simultaneously, V ions in one of the top surface layers show metallicity. It is evident from our analysis that the surface layers of V-doped $Mg_{1-\delta}Ti_2O_4$ have all the key ingredients that can boost superconductivity as the Mott transition gets inhibited. Also, our results are in line with the existing literature that reports (i) reduced Mg/Ti ratio [41] and (ii) suppression of orbital degrees of freedom [23] promoting superconductivity in engineered MgTi$_2$O$_4$ thin films. However, pointing out the primary mechanism for superconductivity from a first principles calculation is extremely challenging for this class of materials and many body approaches are needed. These results will provide a guidance for experimental as well as theoretical explorations of superconductivity in various promising strongly correlated systems.

## Acknowledgements

DD acknowledges Aminur Rahaman and AT acknowledges Debraj Choudhury for very useful discussions on the spinels and their superconductivity.

# Supplemental Materials for
# Possible routes to superconductivity in the surface layers of V-doped $Mg_{1-\delta}Ti_2O_4$ through multiple charge transfers and suppression of Jahn-Teller activity

Dibyendu Dey[1], T. Maitra[2], and A. Taraphder[3]

[1]Department of Physics and Astronomy, University of Maine, Orono, Maine 04469, USA

[2]Department of Physics, Indian Institute of Technology Roorkee, Roorkee, Uttarakhand 247667, India

[3]Department of Physics, Indian Institute of Technology Kharagpur, Kharagpur 721302, India

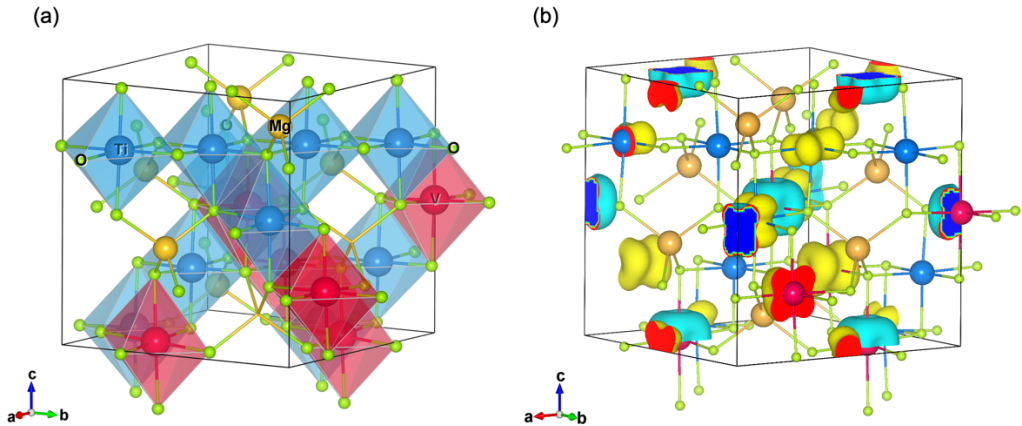

**Figure S1**: (a) Crystal structure of 37.5% bulk V-doped $MgTi_2O_4$. $TiO_6$ and $VO_6$ edge-sharing octahedra are shaded in sky-blue and red colors, respectively. (b) Charge density plot for Ti and V d-states of the same compound. Charge densities depicted in yellow and cyan colors are corresponding to majority and minority spin channels, respectively.

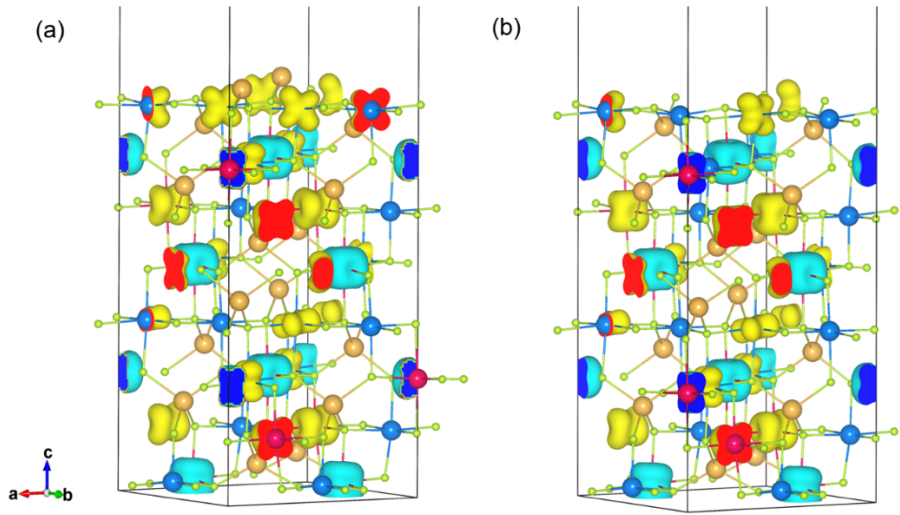

**Figure S2**: Charge density plots for Ti and V d-states of (a) V-doped $Mg_{1-\delta}Ti_2O_4$ and (b) V-doped $Mg_{1-\delta}Ti_2O_4$ slab structures, respectively. Charge densities depicted in yellow and cyan colors are corresponding to majority and minority spin channels, respectively.

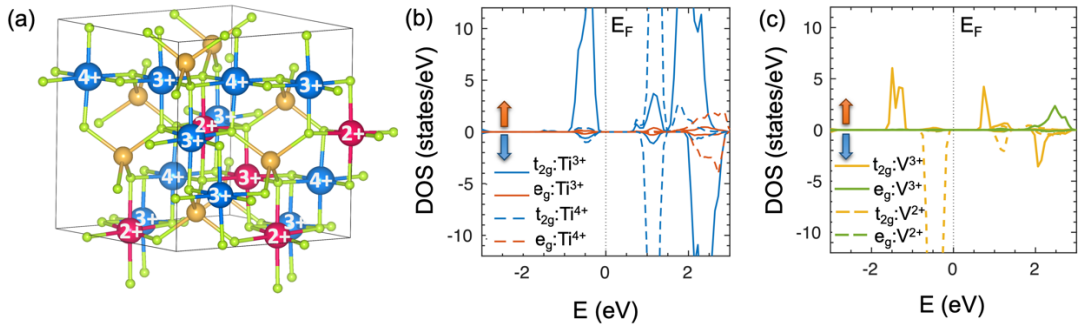

**Figure S3:** (a) Crystal structure of 31.25% bulk V-doped MgTi$_2$O$_4$ compound containing both orbitally inactive Ti$^{4+}$(3d$^0$) and V$^{2+}$(3d$^3$) ions along with orbitally active Ti$^{3+}$ (3d$^1$) and V$^{3+}$ (3d$^2$) ions. $d$-orbital ($t_{2g}$ and $e_g$) projected density of states (DOS) of (b) Ti and (c) V atoms in the majority and minority spin channels as obtained within GGA+$U$ calculations. The results indicate the insulating ground state of the 31.25% bulk V-doped MgTi$_2$O$_4$ compound, where the Fermi-level is denoted as E$_F$. Mg atoms are depicted in yellow, Ti are in blue, V are in red, and O are in green, whereas majority and minority spins are marked as orange and blue arrows, respectively.

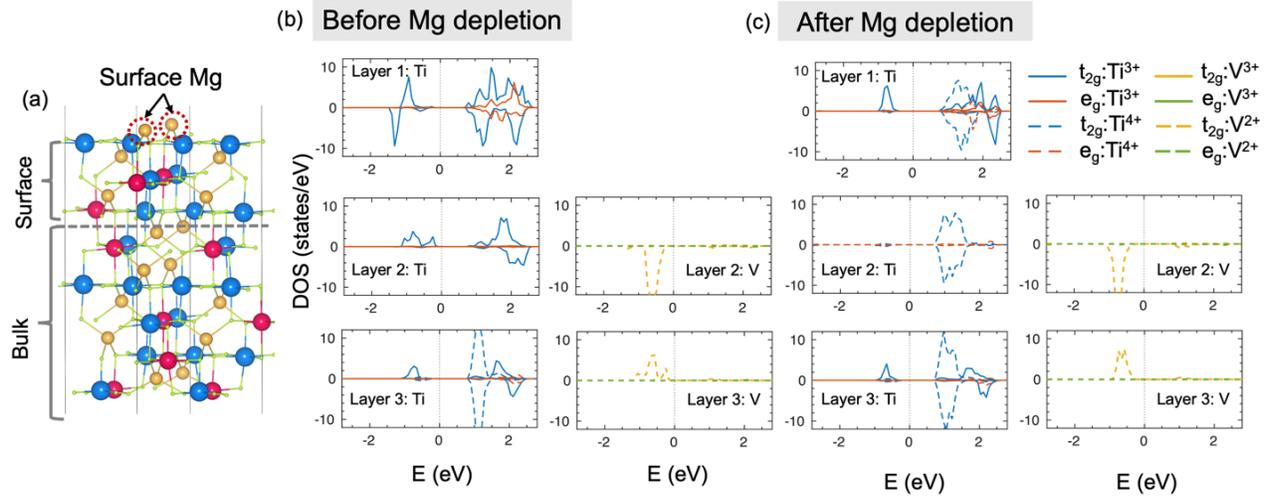

**Figure S4:** (a) V-doped MgTi$_2$O$_4$ slab structure containing 25% V atoms at the surface layers. $d$-orbital ($t_{2g}$ and $e_g$) projected density of states (DOS) of Ti and V atoms in the majority and minority spin channels of each surface layer (b) before and (c) after Mg depletion as obtained within GGA+$U$ calculations.

**Supplemental Material S1: CONTCAR files of 37.5% V-doped bulk and slab structures.**

```
#####################################################################
# CONTCAR for 37.5% V-doped bulk
#####################################################################
CONTCAR
  1.00000000000000
     8.6376475589051509   -0.0152432607466018   -0.0114564806566749
    -0.0151764199228779    8.6681412533501074    0.0049802714527703
    -0.0111306100777945    0.0048433886100030    8.6347518491898576
   Mg   Ti   V    O
    8   10    6   32
Direct
  0.1279451886311733  0.1249564161987706  0.1209545870602966
  0.6279451886311733  0.6249564161987706  0.1209545870602966
  0.3718384907072263  0.3744871526669158  0.8790493531034400
  0.8718384907072263  0.8744871526669158  0.8790493531034400
  0.6210343783616494  0.1239907956751125  0.6202985014803843
  0.1210343783616494  0.6239907956751125  0.6202985014803843
  0.3775252700341412  0.8754191304471988  0.3784401054786031
  0.8775252700341412  0.3754191304471988  0.3784401054786031
  0.4967491809871376  0.7486082280388828  0.7528069467288674
  0.9967491809871447  0.2486082280388828  0.7528069467288674
  0.5054074361272640  0.2535863061720747  0.2468005008720695
  0.0054074361272640  0.7535863061720747  0.2468005008720695
  0.2456088361278290  0.9971205972157691  0.7465798895194808
  0.7456088361278290  0.4971205972157691  0.7465798895194808
  0.5013711652910828  0.0002452266269515  0.0013635863245156
  0.0013711652910828  0.5002452266269515  0.0013635863245156
  0.2498897162561704  0.2522291428327321  0.4989191796928836
  0.7498897162561846  0.7522291428327321  0.4989191796928836
  0.0015315248041787  0.9985370963524360  0.4997041885596332
  0.5015315248041787  0.4985370963524360  0.4997041885596332
  0.2481946119329024  0.7506720131983897  0.0005248899244421
  0.7481946119329024  0.2506720131983897  0.0005248899244421
  0.7513985831940531  0.9995506913579817  0.2534835520612404
  0.2513985831940531  0.4995506913579817  0.2534835520612404
  0.7647938147498294  0.7615577994542733  0.2535191421937668
  0.2647938147498294  0.2615577994542733  0.2535191421937668
  0.2446870885480479  0.2344388001442610  0.7437642950054695
  0.7446870885480479  0.7344388151442587  0.7437642950054695
  0.4834753831479901  0.9946936991838768  0.7553603740211443
  0.9834753831479901  0.4946936991838768  0.7553603740211443
  0.5054305720065528  0.0167107635645323  0.2461885319446964
  0.0054305720065528  0.5167107505645419  0.2461885319446964
  0.2494621358744027  0.0065555251822857  0.5161052407074678
  0.7494621508744004  0.5065555261822894  0.5161052407074678
  0.7546296208411221  0.9940805261552796  0.4872603070832042
  0.2546296208411221  0.4940805261552796  0.4872603070832042
  0.5007847419592153  0.2446430391546386  0.0176306533520929
  0.0007847419592153  0.7446430391546386  0.0176306533520929
  0.4968671937252722  0.7561800645222050  0.9869638031069812
  0.9968671937252651  0.2561800645222050  0.9869638031069812
  0.2518351495530595  0.7649497629385991  0.7496277911822204
  0.7518351495530595  0.2649497629386062  0.7496277911822204
  0.7380211626633582  0.2386120884606555  0.2512505111064911
  0.2380211626633582  0.7386121034606532  0.2512505111064911
  0.4973522355495916  0.4841137563254492  0.2505449937068960
  0.9973522355495916  0.9841137563254492  0.2505449937068960
  0.5137161707071130  0.5037013520515927  0.7506282197663694
  0.0137161707071201  0.0037013520515927  0.7506282197663694
  0.7362049059003866  0.0027379467336033  0.0138895407250601
  0.2362049059003866  0.5027379477336069  0.0138895407250601
  0.2630659909000883  0.9983692551128556  0.9835699136711256
  0.7630659909000883  0.4983692551128556  0.9835699136711256
  0.5146003006100486  0.7474542350926328  0.5123683706432445
  0.0146003006100486  0.2474542350926328  0.5123683706432445
  0.4865791358091016  0.2517985701400534  0.4824030249779199
  0.9865791358091016  0.7517985701400463  0.4824030249779199
```

```
##########################################################################
# CONTCAR for 37.5% V-doped slab (before Mg depletion)
##########################################################################
CONTCAR
   1.000000000000
     8.6376686096000004    0.0000000000000000    0.0000000000000000
    -0.0304798141000000    8.6681020820000008    0.0000000000000000
    -0.0451853678000000    0.0194906232000000   34.2694501419000019
   Mg  Ti  V   O
   16  20  12  64
Direct
 0.1320481352693719  0.1263333710575125  0.0386878466298981
 0.1271771436754747  0.1259924302504984  0.2953747716642710
 0.6320481202693884  0.6263333560575006  0.0386878466298981
 0.6271771586754582  0.6259924742504950  0.2953747716642710
 0.3715695739337974  0.3745095974777826  0.2341386793836548
 0.3642738912174295  0.3805451589778173  0.4799897492821401
 0.8715696039337786  0.8745096274777922  0.2341386793836548
 0.8642739502174237  0.8805451289778290  0.4799897492821401
 0.6194139420467621  0.1244581137183971  0.1684773704762961
 0.6200155676463694  0.1231052377452571  0.4229201083781078
 0.1194139500467628  0.6244581277183698  0.1684773704762961
 0.1200155526463789  0.6231051857452456  0.4229201083781078
 0.3729991762364193  0.8719181618771259  0.1089198116873007
 0.3748333945071565  0.8780491791308478  0.3589591914868180
 0.8729992062364360  0.3719181918771071  0.1089198116873007
 0.8748333945071423  0.3780491491308382  0.3589591914868180
 0.4963355015659445  0.7486474366165368  0.2023463238771797
 0.4891030125457476  0.7572359445722014  0.4509307603025050
 0.9963355605659316  0.2486474366165439  0.2023463238771797
 0.9891029825457309  0.2572359445721872  0.4509307603025050
 0.5002805924534925  0.2637794742576389  0.0735875999537399
 0.5042899002950563  0.2548631465845901  0.3273816604699817
 0.0002805554534859  0.7637794742576460  0.0735875999537399
 0.0042899042950566  0.7548631465845546  0.3273816604699817
 0.2449319088715569  0.9971957115216981  0.2009056207428515
 0.2479877781308701  0.9921970437611591  0.4513512742363019
 0.7449318488715733  0.4971956525216967  0.2009056207428515
 0.7479877631308653  0.4921969537611233  0.4513512742363019
 0.4975057885015417  0.9917290070673488  0.0189260496477246
 0.5013837656691109  0.0015369661463112  0.2647944427306541
 0.9975058405015389  0.4917290200673392  0.0189260496477246
 0.0013837406691124  0.5015369751463226  0.2647944427306541
 0.2491279537892481  0.2526689051068018  0.1378715506368664
 0.2476880244511150  0.2503687541848407  0.3902025989255833
 0.7491279977892376  0.7526688751067780  0.1378715506368664
 0.7476880094511031  0.7503687841848503  0.3902025989255833
 0.0000432001927635  0.9990495593345798  0.1378029570520241
 0.9983638005989661  0.9999579621304306  0.3902537212109110
 0.5000431891927590  0.4990495293345560  0.1378029570520241
 0.4983638305989615  0.4999579621305159  0.3902537212109110
 0.2549823266475997  0.7517803234675497  0.0156173988124664
 0.2473247580917288  0.7511228843679660  0.2643782267086578
 0.7549823716476212  0.2517802644675768  0.0156173988124664
 0.7473247580917359  0.2511229143679472  0.2643782267086578
 0.7509075361504500  0.9992071166141301  0.0756768979133753
 0.7500086233130148  0.0007344057900838  0.3287821035096670
 0.2509075361504358  0.4992071466141539  0.0756768979133753
 0.2500086533130172  0.5007344057901193  0.3287821035096670
 0.7646038872356513  0.7618750781860086  0.0754653787545791
 0.7645820341980780  0.7625732410663346  0.3285631952633778
 0.2646038582357022  0.2618750181860179  0.0754653787545791
 0.2645820641980947  0.2625731810663368  0.3285631952633778
 0.2445530932026330  0.2348046085674440  0.1995846328158564
 0.2408039115656564  0.2306980717188551  0.4538368734194833
 0.7445530482026683  0.7348046675674382  0.1995846328158564
 0.7408038815656539  0.7306980567188432  0.4538368734194833
 0.4825338309621117  0.9950320287209493  0.2026465227064946
 0.4912617586560444  0.9815113218428237  0.4564935246992263
 0.9825338009620666  0.4950320287209706  0.2026465227064946
```

```
0.9912616986560394  0.4815112918428426  0.4564935246992263
0.5034196448768569  0.0111751849596828  0.0751986636977406
0.5049302465903480  0.0184480611868807  0.3262149296715222
0.0034196358768597  0.5111751699596994  0.0751986636977406
0.0049302185903670  0.5184480431868792  0.3262149296715222
0.2477678974320270  0.0076792300710835  0.1426095009134798
0.2445517112700770  0.0086639347782977  0.3920867091662217
0.7477678834320258  0.5076791720711213  0.1426095009134798
0.7445517112700770  0.5086639337783083  0.3920867091662217
0.7527391880942744  0.9944065273645251  0.1347279983477918
0.7510776463520870  0.9938894861173821  0.3878590831194657
0.2527391880942886  0.4944065573645204  0.1347279983477918
0.2510776173520455  0.4938894861173537  0.3878590831194657
0.5131730474296745  0.2273894416250286  0.0181115294268110
0.5001343543427055  0.2452567161457964  0.2692407799361760
0.0131730324296342  0.7273894416250215  0.0181115294268110
0.0001342953427184  0.7452567161457893  0.2692407799361760
0.4966198286949606  0.7566381835190015  0.2611686241828934
0.4894350878294063  0.7709810319545483  0.0127784270132736
0.9966197986949439  0.2566381835190015  0.2611686241828934
0.9894351468294360  0.2709810619545507  0.0127784270132736
0.2511339757723974  0.7650635254520637  0.2010012615686065
0.2501793811400930  0.7656509617340248  0.4560429526641698
0.7511339757723903  0.2650635254520850  0.2010012615686065
0.7501794411400837  0.2656510217340440  0.4560429526641698
0.7344091233355527  0.2392931932105995  0.0760925113739930
0.7364277992599355  0.2390768284703242  0.3274332266207054
0.2344091533355481  0.7392931782105876  0.0760925113739930
0.2364278732599274  0.7390768734702959  0.3274332266207054
0.4975521495832282  0.4890092235759482  0.0742392034226356
0.4956752139726817  0.4846866548139559  0.3276148728292085
0.9975521795832165  0.9890092535759862  0.0742392034226356
0.9956752439726841  0.9846866248139534  0.3276148728292085
0.5129895920131773  0.5034436341681143  0.2011681576083575
0.5014480836918693  0.5216663256462581  0.4554958692574758
0.0129896190131760  0.0034436401681219  0.2011681576083575
0.0014480916918700  0.0216663346462553  0.4554958692574758
0.7253738015506599  0.0166504188144359  0.0173020738017158
0.7357846764484393  0.0031073207249719  0.2680535610289922
0.2253738015506741  0.5166504258144613  0.0173020738017158
0.2357846614484416  0.5031073247249935  0.2680535610289922
0.2618971898539755  0.9994352146665690  0.2602312125361692
0.2749623353601223  0.9859260692812057  0.0111930166932410
0.7618971898539826  0.4994351256665723  0.2602312125361692
0.7749623053601340  0.4859260102811618  0.0111930166932410
0.5136617675179096  0.7471899030846245  0.1417197601459605
0.5134504387091212  0.7476205352390579  0.3910314998198317
0.0136618015179621  0.2471899480846531  0.1417197601459605
0.0134504547091367  0.2476205202390460  0.3910314998198317
0.4860414101135220  0.2518901591583429  0.1331653335670921
0.4858236991437295  0.2517270363756765  0.3871405312084875
0.9860413801135337  0.7518901881583844  0.1331653335670921
0.9858236691437128  0.7517270063756598  0.3871405312084875
```

###########################################################################
# CONTCAR for 37.5% V-doped slab (*after* Mg depletion)
###########################################################################
CONTCAR
   1.000000000000
     8.6376686096000004    0.0000000000000000    0.0000000000000000
    -0.0304798141000000    8.6681020820000008    0.0000000000000000
    -0.0451853678000000    0.0194906232000000   34.2694501419000019
   Mg   Ti   V   O
   14   20   12   64
Direct
  0.1314730997659908  0.1252886998202456  0.0388611013183464
  0.1263269442119892  0.1245495474178995  0.2965712543292511
  0.6314730847660073  0.6252886848202479  0.0388611013183464
  0.6263269592120011  0.6245495914178960  0.2965712543292511
  0.3709030447484665  0.3744953222441438  0.2336584889903932
  0.8709030747484618  0.8744953522441463  0.2336584889903932
  0.6187277543645564  0.1239307423933269  0.1689762109509587
  0.6150014446116572  0.1248873969221407  0.4262865396324571
  0.1187277623645500  0.6239307563933636  0.1689762109509587
  0.1150014296116453  0.6248873449221435  0.4262865396324571
  0.3720050460434692  0.8721531726577254  0.1093753349566100
  0.3762021105795057  0.8760593392381182  0.3583710220753886
  0.8720050760434574  0.3721532026577421  0.1093753349566100
  0.8762021105795057  0.3760593092380944  0.3583710220753886
  0.4958945624924738  0.7482676284966203  0.2029298377104993
  0.4952178261787452  0.7510676945331340  0.4507391058616008
  0.9958946214924751  0.2482676284966203  0.2029298377104993
  0.9952177961787712  0.2510676945331269  0.4507391058616008
  0.5001465117643917  0.2637383665253878  0.0741570446183886
  0.5032518675337769  0.2533143245845153  0.3275000909341017
  0.0001464747643851  0.7637383665253950  0.0741570446183886
  0.0032518715337702  0.7533143245845082  0.3275000909341017
  0.2449767734946562  0.9963657016829970  0.2014604512540075
  0.2484366466407124  0.9936170145018792  0.4472979476123768
  0.7449767134946512  0.4963656426830028  0.2014604512540075
  0.7484366316407360  0.4936169245018647  0.4472979476123768
  0.4982414474022079  0.9925576498095197  0.0191596297908490
  0.5023671030403065  0.0016660514836033  0.2648280842012483
  0.9982414994022193  0.4925576628095243  0.0191596297908490
  0.0023670780403151  0.5016660604835934  0.2648280842012483
  0.2472506542162165  0.2523284560332399  0.1378170648275443
  0.2512722546450590  0.2645240881581756  0.3920203805446363
  0.7472506982162130  0.7523284260332446  0.1378170648275443
  0.7512722396450471  0.7645241181581639  0.3920203805446363
  0.9996776991814329  0.9986354639869788  0.1376300640683255
  0.9970369250786888  0.9997309218531782  0.3910218296732282
  0.4996776881814284  0.4986354339869763  0.1376300640683255
  0.4970369550786913  0.4997309218531711  0.3910218296732282
  0.2552313175455296  0.7504412263241917  0.0158711231118929
  0.2460581198498559  0.7516267225793740  0.2643320650690200
  0.7552313625455227  0.2504411673242046  0.0158711231118929
  0.7460581198498488  0.2516267525793694  0.2643320650690200
  0.7506885487599249  0.9975839625073135  0.0762461336262135
  0.7483985175637073  0.0014311277972112  0.3280238514229126
  0.2506885487599320  0.4975839925073231  0.0762461336262135
  0.2483985475637169  0.5014311277972183  0.3280238514229126
  0.7648264592835687  0.7603437715362418  0.0752412427022193
  0.7625288661839491  0.7553306520592429  0.3316641269904537
  0.2648264302835699  0.2603437115362510  0.0752412427022193
  0.2625288961839303  0.2553305920592379  0.3316641269904537
  0.2441410057313433  0.2343981507509056  0.1989144226169941
  0.2434265707170979  0.2289412476452171  0.4475702223432663
  0.7441409607313432  0.7343982097508999  0.1989144226169941
  0.7434265407171026  0.7289412326452194  0.4475702223432663
  0.4809859531660337  0.9944134730287360  0.2023050781377904
  0.4742871618722120  0.9836287197482392  0.4556037775987960
  0.9809859231660383  0.4944134730287288  0.2023050781377904
  0.9742871018721999  0.4836286897482367  0.4556037775987960
  0.5032484123656360  0.0106459434096351  0.0752358999960876

```
0.5025044965532430  0.0173897237233618  0.3259580988078099
0.0032484033656317  0.5106459284096374  0.0752358999960876
0.0025044685532407  0.5173897057233887  0.3259580988078099
0.2464509657384681  0.0087317374192608  0.1424174286246114
0.2453879752752854  0.0137649644130136  0.3913244996562000
0.7464509517384599  0.5087316794192560  0.1424174286246114
0.7453879752752925  0.5137649634130241  0.3913244996562000
0.7520666870392176  0.9942255580220234  0.1350054474205322
0.7486103811084135  0.9889563009505338  0.3911243443896169
0.2520666870392105  0.4942255580220188  0.1350054474205322
0.2486103521084289  0.4889563009505338  0.3911243443896169
0.5152806505175889  0.2274766717892476  0.0180780910876379
0.4974740324603530  0.2458528775495452  0.2692169174678014
0.0152806355176196  0.7274766717892476  0.0180780910876379
0.9974739734603517  0.7458528775495452  0.2692169174678014
0.4968425523865392  0.7556226783858619  0.2614952076852219
0.4879953926730067  0.7705279379556558  0.0130892111155489
0.9968425223865438  0.2556226783858691  0.2614952076852219
0.9879954516730010  0.2705279679556369  0.0130892111155489
0.2510482513039065  0.7649922968091616  0.2006567558879979
0.2561250771869510  0.7725180030679795  0.4522509410618412
0.7510482513038923  0.2649922968091687  0.2006567558879979
0.7561251371869631  0.2725180630679915  0.4522509410618412
0.7347616899398091  0.2389837497778444  0.0759001482736110
0.7333848696186251  0.2460534320085088  0.3272051480489537
0.2347617199398044  0.7389837347778538  0.0759001482736110
0.2333849436186171  0.7460534770085232  0.3272051480489537
0.4972124081154234  0.4886891653718735  0.0741181305357017
0.4935589307649622  0.4839099322770011  0.3285299989942132
0.9972124381154117  0.9886891953718759  0.0741181305357017
0.9935589607649504  0.9839099022769986  0.3285299989942132
0.5129953202331308  0.5031630844536039  0.2006050006519331
0.5258494454878431  0.5232439480199886  0.4543073035029721
0.0129953472331437  0.0031630904536044  0.2006050006519331
0.0258494534878579  0.0232439570199929  0.4543073035029721
0.7260716421405675  0.0176056597682148  0.0171337711018609
0.7336410645850933  0.0051098997772812  0.2653434461160487
0.2260716421405533  0.5176056667682190  0.0171337711018609
0.2336410495850885  0.5051099037772815  0.2653434461160487
0.2603810971303417  0.0006869127679892  0.2601439276236590
0.2746931779065918  0.9834571156357086  0.0115872297537436
0.7603810971303560  0.5006868237680138  0.2601439276236590
0.7746931479065893  0.4834570566357002  0.0115872297537436
0.5134442534869024  0.7463886097466386  0.1414941261607154
0.5173026981639453  0.7469861202922843  0.3904761517328836
0.0134442874869123  0.2463886547466387  0.1414941261607154
0.0173027141639466  0.2469861052922795  0.3904761517328836
0.4855548135136374  0.2514452241139935  0.1330928099405497
0.4853853516370208  0.2521379241742636  0.3875103283924801
0.9855547835136349  0.7514452531139995  0.1330928099405497
0.9853853216370041  0.7521378941742682  0.3875103283924801
```

**Supplemental Material S2: CONTCAR files of 31.5% V-doped bulk structures for different Ti and V atomic arrangements.**

```
####################################################################
# CONTCAR 1 for 31.5% V-doped bulk
####################################################################
CONTCAR 1
  1.00000000000000
     8.6376475589051509   -0.0152432607466018   -0.0114564806566749
    -0.0151764199228779    8.6681412533501074    0.0049802714527703
    -0.0111306100777945    0.0048433886100030    8.6347518491898576
   Mg   Ti   V    O
    8   11    5   32
Direct
  0.1272995218842894  0.1240542454092335  0.1212374905334386
  0.6278106191152375  0.6254347408402836  0.1234518923136960
  0.3698810459667854  0.3757117960135332  0.8769607579433725
  0.8715985709227780  0.8751861821720297  0.8784050974392912
  0.6217114446556025  0.1257064852946499  0.6223297987855574
  0.1240090575638249  0.6233606010771666  0.6202978717993872
  0.3772930369548675  0.8750514220153605  0.3777729043893956
  0.8757701890643474  0.3768425363165520  0.3773500018300027
  0.4970431829113195  0.7492171773914436  0.7546669703262125
  0.9973082938795841  0.2492934105589555  0.7539358047647013
  0.5050340688878450  0.2511650453686727  0.2436100969202997
  0.0057323885685534  0.7524358940165357  0.2451524299206085
  0.2456791757671013  0.9968343321281523  0.7476735640519578
  0.7483365615708237  0.4981384063776702  0.7514919063625314
  0.5019092287771372  0.0003585477955070  0.0008396686391237
  0.9987570251660287  0.4993119486053033  0.0000970422148541
  0.2475992684888126  0.2479046405739069  0.4981291828816907
  0.7524800796893700  0.7554486685027442  0.4990813047381053
  0.5023661183315440  0.4993727323469699  0.5005688593399142
  0.0015035579355001  0.9988000237160506  0.5000486774692163
  0.2483024104055787  0.7505825054176469  0.0004060171009499
  0.7474227406361180  0.2504504334774964  0.0025559408027458
  0.7518041628542775  0.9992249561279252  0.2540087687116497
  0.2508705150312593  0.4987893944640689  0.2489712539575635
  0.7657178464283732  0.7624642625225988  0.2540781984737919
  0.2645625777929439  0.2604538809895374  0.2532340144638638
  0.2450443163305636  0.2329005597124052  0.7426699930799572
  0.7443797570398871  0.7379668436877509  0.7430366605370722
  0.4836616561113161  0.9953732482166373  0.7554691594072338
  0.9943973311787815  0.4929770123446815  0.7555736717937123
  0.5048127031672820  0.0164731490377719  0.2463840391824164
  0.0080274874181399  0.5161072930767361  0.2475256507672512
  0.2496198136109058  0.0052466914794849  0.5156840860347174
  0.7360144824613073  0.5077877914926745  0.5034356187864617
  0.7546244305727683  0.9958564991391867  0.4882348866656585
  0.2645733103188164  0.4945044146961450  0.4864916914689772
  0.4995381148191385  0.2452776820302773  0.0157258960537092
  0.0012507833725266  0.7451368081337009  0.0168956485783411
  0.4970052324409053  0.7564651610995341  0.9886583293292617
  0.9959233120562914  0.2545058498855539  0.9884596953753046
  0.2526169699040111  0.7655313133627288  0.7493473843066880
  0.7529156873063911  0.2624984426967032  0.7519383132458088
  0.7371567104836458  0.2376610439316948  0.2523416397247402
  0.2384684840158968  0.7381206784095369  0.2509681792489289
  0.4955587653915856  0.4838523502133896  0.2521319691275039
  0.9976718856196740  0.9832131222119642  0.2513041087042893
  0.5018330818507195  0.5063006266555306  0.7442303568110944
  0.0144392280478627  0.0034859374360821  0.7506234574415629
  0.7358717150659473  0.0025716144820009  0.0144885967445134
  0.2361737815497520  0.5028925764440615  0.0146738232594643
  0.2633697489806934  0.9980897273426095  0.9840088674340421
  0.7646297825168915  0.4971968082730243  0.9960965764642538
  0.5144937651152190  0.7454727519361981  0.5106904477732996
  0.0145788578940156  0.2491288239139209  0.5110127767016692
  0.4860230581500105  0.2549766939423250  0.4829409980701982
  0.9875230429591682  0.7508341961957186  0.4825719497079390
```

###########################################################################
# **CONTCAR 2** for 31.5% V-doped bulk
###########################################################################
CONTCAR 2
   1.00000000000000
     8.6376475589051509   -0.0152432607466018   -0.0114564806566749
    -0.0151764199228779    8.6681412533501074    0.0049802714527703
    -0.0111306100777945    0.0048433886100030    8.6347518491898576
    Mg   Ti   V   O
     8   11   5   32
Direct
  0.1245120664205004  0.1258979937029849  0.1209961185587289
  0.6273474499036951  0.6238692236919050  0.1233488031580023
  0.3738974215856814  0.3730069652985932  0.8787013046687804
  0.8720092483255968  0.8748305715169877  0.8784700307547197
  0.6217616988573553  0.1232823244195771  0.6231779191633606
  0.1215857810930032  0.6246642764739647  0.6203403168592843
  0.3780322587134606  0.8747462499503982  0.3777087287550073
  0.8791048591063344  0.3744015161551104  0.3763688105293497
  0.4964452697421819  0.7481792876076128  0.7529798961308458
  0.9988146026327200  0.2494952938038466  0.7473186682262400
  0.5018111644206655  0.2525846665561531  0.2517268285094900
  0.0054547752839795  0.7538248058796100  0.2476210250173239
  0.2453443364868022  0.9978097876378200  0.7448758431162474
  0.7455786722606774  0.4992669366688531  0.7435747247645779
  0.4981460052453812  0.9971894416807956  0.0016637250080294
  0.0032837912824490  0.5036624206017564  0.0003654768550803
  0.2532867024065268  0.2530080758683653  0.4977410192169245
  0.7495741737965176  0.7522019074788133  0.4984853446216135
  0.7471104782602467  0.2498519218603619  0.0014273511186929
  0.0012137001467210  0.9986257501848854  0.4996012310562961
  0.5020152941977472  0.4989057442583089  0.5014850493510252
  0.2483296175756209  0.7507843730545503  0.0006369289412547
  0.7514733376412437  0.0004163470445278  0.2545413333437194
  0.2504925889547494  0.4988758047136841  0.2541622952330016
  0.7650602671577431  0.7618790310143524  0.2535550279213581
  0.2537035049794270  0.2631903143245893  0.2538052695834239
  0.2434622668900772  0.2345397063365553  0.7456122615391649
  0.7452638514803667  0.7345189691124787  0.7442464462831495
  0.4825042330644962  0.9931020478067936  0.7560269602481711
  0.9837933098501068  0.4952740412932428  0.7549567229169867
  0.5066770931348543  0.0130535443202220  0.2457554867535308
  0.0045844293836197  0.5183122457473530  0.2448568121470842
  0.2490928993189883  0.0061144618679378  0.5153709425697173
  0.7505116197278880  0.5057109408969964  0.5146308673195747
  0.7543736777924863  0.9936112831481339  0.4885079165231758
  0.2547782571416377  0.4951291796362582  0.4879143561576456
  0.5142831717569081  0.2436607524848071  0.0057808220367690
  0.0007889003284944  0.7458103143845136  0.0170891064238674
  0.4970474376551053  0.7544407580854227  0.9875032683151659
  0.9859417299847806  0.2552215980697099  0.9862201059474742
  0.2517222160438024  0.7654139695072786  0.7501213108279075
  0.7535021641990980  0.2652163923405269  0.7526714118906312
  0.7498002254885563  0.2359424700613744  0.2457328076304250
  0.2372748275750354  0.7392313663833505  0.2508102224242350
  0.4954541825504393  0.4871965067281963  0.2524478363578382
  0.9967162707445709  0.9836413360474552  0.2506324334262757
  0.5144416213175091  0.5042605340811477  0.7518156922333645
  0.0133810122223679  0.0045781831741891  0.7504940891075691
  0.7360856661872859  0.0051238799813049  0.0118835340441947
  0.2361263738853889  0.5014709692690218  0.0123708195999797
  0.2617798036518408  0.9997535423052213  0.9837427801275211
  0.7635521923421109  0.4953025972019631  0.9845704245336648
  0.5148756372345460  0.7477444383018792  0.5131156790160887
  0.0144322601608593  0.2475881110236813  0.5128190111237174
  0.4858062888547181  0.2526754569056351  0.4949166437797174
  0.9865572990550007  0.7519093530489513  0.4827048962430496

###########################################################################
# **CONTCAR 3** for 31.5% V-doped bulk
###########################################################################
CONTCAR 3
  1.000000000000000
     8.6376475589051509   -0.0152432607466018   -0.0114564806566749
    -0.0151764199228779    8.6681412533501074    0.0049802714527703
    -0.0111306100777945    0.0048433886100030    8.6347518491898576
   Mg   Ti   V   O
    8   11   5   32
Direct
 0.1277591316991078  0.1239202831063366  0.1210100647291625
 0.6290973019245527  0.6264352533954494  0.1213215998594848
 0.3721109925185928  0.3737558923344153  0.8786386048731529
 0.8714729774864765  0.8744605281294184  0.8787349592809335
 0.6217092525833792  0.1236369235413761  0.6206566355757346
 0.1207692754385619  0.6244398346706390  0.6209892802103738
 0.3787673517855055  0.8772421888946198  0.3790522923801376
 0.8769292760007090  0.3748025770860011  0.3777000347272761
 0.4968600980425606  0.7486333035285639  0.7529571975397999
 0.9964837999121698  0.2481287727712385  0.7525362215996552
 0.5035937910895640  0.2520988582334240  0.2473901718366136
 0.0062153690223354  0.7548691484014611  0.2466831369790867
 0.2454458341525481  0.9968333167764598  0.7474392376509371
 0.7453431838037048  0.4968856791193090  0.7462597772258448
 0.5017964916516746  0.0004382479784084  0.0013134816296585
 0.9991813583355409  0.5004571380342142  0.9999893164626741
 0.2502090651580957  0.2492478706724270  0.5002603764155182
 0.7501207924811197  0.7521069873050976  0.4988316050629891
 0.2534352552231098  0.5014489220748644  0.2528316403398421
 0.0012816470157944  0.9984879403138436  0.4995304711742250
 0.5021409913795480  0.4982224358597946  0.5004891245178058
 0.2478277136720095  0.7514982994119563  0.0001986748655440
 0.7484140308872753  0.2510171423118237  0.0005378312028199
 0.7515301789541624  0.0003207684584581  0.2531124368451714
 0.7647847510695200  0.7627278278137553  0.2538192030774127
 0.2666587715710165  0.2545959869294734  0.2545662011188128
 0.2445498439886293  0.2339379937274444  0.7449737025898244
 0.7446767994711081  0.7341312369067268  0.7438188629841278
 0.4835474050835202  0.9945798431486139  0.7554345162339544
 0.9832392242993606  0.4945504279706086  0.7548865705140315
 0.5070387183715539  0.0160691066814920  0.2459624858037230
 0.0041483090710059  0.5169955599817726  0.2450860737542015
 0.2496744143754484  0.0055593186928107  0.5172154485136105
 0.7493149753620187  0.5061613976883876  0.5159962259854254
 0.7544307006254840  0.9936571024111842  0.4873060614236024
 0.2528826772252799  0.4935459902207384  0.4868167306340609
 0.5009623133959877  0.2448509854965408  0.0178708523209963
 0.0001315650479086  0.7443672849555512  0.0170783197349209
 0.4966506182860897  0.7562152692212649  0.9870888453376239
 0.9975752947081205  0.2568136293367616  0.9869696112825892
 0.2519726174767456  0.7645225569156082  0.7498864370625000
 0.7517535595585585  0.2649701705228225  0.7497816334331588
 0.7376304642010467  0.2382155334444889  0.2509146872735002
 0.2370024090335647  0.7473325383842990  0.2507260068089536
 0.4996359834242980  0.4844519943637522  0.2503488441278492
 0.9956698473924206  0.9851426766789189  0.2502722353013382
 0.5133710318720262  0.5034901301454227  0.7504152773142110
 0.0136019986723639  0.0033638232820650  0.7506365609071040
 0.7362898974154390  0.0030618945227872  0.0130881048853979
 0.2361574677347704  0.5024295322620205  0.0136363505517494
 0.2632511830698192  0.9982859494392500  0.9832851827226889
 0.7619860281785122  0.4990267251953640  0.9831023204648659
 0.5143924260940551  0.7471623381107833  0.5126523574797659
 0.0140574609821584  0.2476822697521328  0.5127314758513819
 0.4875478176256394  0.2513213890286394  0.4829874033470247
 0.9869182500984408  0.7513931853589710  0.4821812261751646

**Table S1:** Total energy and number of Jahn-Teller (JT) active ($Ti^{3+}/V^{3+}$) and JT inactive ($Ti^{4+}$, $V^{2+}$) ions of 31.5% V-doped bulk structures for different Ti and V atomic arrangements.

| Structures | Energy (eV) | $Ti^{3+}$ | $Ti^{4+}$ | $V^{3+}$ | $V^{2+}$ |
|---|---|---|---|---|---|
| CONTCAR 1 | -414.31142 | 8 | 3 | 2 | 3 |
| CONTCAR 2 | -414.35581 | 8 | 3 | 2 | 3 |
| CONTCAR 3 | -414.53991 | 7 | 4 | 1 | 4 |

## Supplemental Material S3: Length distortion parameter ($\zeta$) calculations

An alternate measure of the JT effect involves analysis of the associated Ti/V-O distortions. To illustrate this, we consider 37.5% V-doped $MgTi_2O_4$ bulk compound to estimate the length distortion parameter ($\zeta$) [1] of 16 octahedra in the supercell structure.

$$\zeta = \sum_{i=1}^{6} |d_i - \langle d \rangle|$$

The parameter $\zeta$ is the sum of the deviation of 6 metal-ligand bond lengths around the central metal atom ($d_i$) from the average value $\langle d \rangle$. As shown in Table S2, $\zeta$ values are smaller for non-JT transition metal sites than JT active sites.

**Table S2:** The length distortion parameter ($\zeta$) of various $TiO_6$ and $VO_6$ octahedra of 37.5% V-doped $MgTi_2O_4$ compound.

| Octahedra | JT active | $\zeta$ (Å) |
|---|---|---|
| $Ti^{4+}$-$O_6$ | N | 0.17 |
| $Ti^{4+}$-$O_6$ | N | 0.17 |
| $Ti^{4+}$-$O_6$ | N | 0.16 |
| $Ti^{4+}$-$O_6$ | N | 0.16 |
| $Ti^{3+}$-$O_6$ | Y | 0.23 |
| $Ti^{3+}$-$O_6$ | Y | 0.23 |
| $Ti^{3+}$-$O_6$ | Y | 0.21 |
| $Ti^{3+}$-$O_6$ | Y | 0.21 |
| $Ti^{3+}$-$O_6$ | Y | 0.21 |
| $Ti^{3+}$-$O_6$ | Y | 0.21 |
| $V^{2+}$-$O_6$ | N | 0.05 |
| $V^{2+}$-$O_6$ | N | 0.05 |
| $V^{2+}$-$O_6$ | N | 0.05 |
| $V^{2+}$-$O_6$ | N | 0.05 |
| $V^{3+}$-$O_6$ | Y | 0.20 |
| $V^{3+}$-$O_6$ | Y | 0.20 |